\documentclass[aps,twocolumn,lengthcheck,10pt,prl,superscriptaddress,amsmath,amssymb]{revtex4-1}

\usepackage{color}
\usepackage{pifont}
\usepackage{hyperref}
\usepackage{graphicx,epstopdf}
\usepackage{textcomp}
\usepackage{tabularx}
\usepackage{booktabs}
\usepackage{array}
\usepackage{bbm}
\usepackage{bm}
\usepackage{amssymb,amsmath}
\usepackage{fix-cm}
\usepackage[all]{xy}
\usepackage{ifpdf}

\usepackage{wasysym}

\providecommand{\ket}[1]{|#1\rangle}
\providecommand{\bra}[1]{\langle#1|}

\providecommand{\idop}{\mathbbm 1}

\newcolumntype{C}[1]{>{\centering\arraybackslash$}p{#1}<{$}}

\begin{document}
\title{Self assembled Wigner crystals as mediators of spin currents and quantum information }

\author{Bobby Antonio}
\affiliation{Department of Physics and Astronomy, University College London, Gower Street, WC1E 6BT London, United Kingdom}

\author{Abolfazl Bayat}
\affiliation{Department of Physics and Astronomy, University College London, Gower Street, WC1E 6BT London, United Kingdom}

\author{Sanjeev Kumar}
\affiliation{Department of Electronic and Electrical Engineering, University College London, London, WC1E 7JE, United Kingdom}
\affiliation{London Centre for Nanotechnology, 17-19 Gordon Street, London, WC1H 0AH, United Kingdom}

\author{Michael Pepper}
\affiliation{Department of Electronic and Electrical Engineering, University College London, London, WC1E 7JE, United Kingdom}
\affiliation{London Centre for Nanotechnology, 17-19 Gordon Street, London, WC1H 0AH, United Kingdom}

\author{Sougato Bose}
\affiliation{Department of Physics and Astronomy, University College London, Gower Street, WC1E 6BT London, United Kingdom}


\begin{abstract}
Technological applications of many-body structures that emerge in gated devices under minimal control are largely unexplored. Here we show how emergent Wigner crystals in a semiconductor quantum wire can facilitate a pivotal requirement for a scalable quantum computer, namely transmitting quantum information encoded in spins faithfully over a distance of micrometers. The fidelity of the transmission is remarkably high, faster than the relevant decohering effects, independent of the details of the spatial charge configuration in the wire, and realizable in dilution refrigerator temperatures. The transfer can evidence near unitary many-body nonequilibrium dynamics hitherto unseen in a solid-state device.  It could also be useful in spintronics as a method for pure spin current over a distance without charge movement.
\end{abstract}

\maketitle
{\em Introduction.--} Spin chains can facilitate several important technological applications such as pure spin currents in non-itinerant systems~\cite{Van2011}, quantum state transfer~\cite{Bose2008,Nikolopoulos2013} and quantum gates~\cite{Yao2011,Banchi2011,Schaffry2012}. Most of the above applications are facilitated by a nearly unitary dynamics of  the spin chain. Such dynamics is not only interesting for potential quantum technology~\cite{Bayat2010kondo,Gu-PRA-2012} but also fundamentally important to address questions of equilibration, quantum thermodynamics and information propagation~\cite{Gogolin2015}. Thus, the physical realization of artificial spin chains that have the potential for long time unitary dynamics is an important quest -- they have been realized only very recently, and exclusively in atomic physics systems: in cold atom systems~\cite{Fukuhara2013,Fukuhara2015}, ion traps~\cite{Richerme2014,Jurcevic2014} and Rydberg systems~\cite{Barredo2015}. In the realm of solid-state, on the other hand, nonequilibrium dynamics of engineered spin chains is far from unitary and primarily driven by equilibration/relaxation~\cite{Ako2011}. Although some bulk magnetic materials~\cite{Sahling2015,Mitra2015}, NV centre chains~\cite{Yao2011,Schaffry2012} and Josephson junction arrays~\cite{Las2014} hold the potential for long chain unitary dynamics, that is still somewhat distant from experimental realization. As far as the semiconductor realm is concerned, permanently fabricated spin chain structures with dangling bonds~\cite{Wolkow2014} or phosphorus dopants ~\cite{Zwanenburg2013} may also hold the potential, but is yet to be examined either theoretically or experimentally. A natural question is thus whether one can realize spin chains exhibiting nearly unitary nonequilibrium dynamics in a feasible manner in a two dimensional electron gas (2DEG) and thereby open the door to the aforementioned applications in solid state.

Individual electrons trapped in gate defined quantum dot arrays have been proposed for simulating spin chains in the limit of one electron in each dot~\cite{Stafford1994}. Although the fabrication of large quantum dot arrays is an active current effort~\cite{Barth2013, Puddy2014} and the complex electronics for gate addressing is also being designed~\cite{Puddy2014}, it is worth considering the potential of simpler gated structures. In particular, self assembled charge configurations, such as Wigner crystals, are naturally formed without demanding local control. By controlling the density of electrons one can vary the distance between the charges and consequently engineer their exchange interactions. This, in comparison with quantum dot arrays, allows for stronger exchange couplings which then potentially provide more thermal stability, faster dynamics and less sensitivity to decoherence.

Here we show that emergent ``self assembled" electronic spin chains arising due to Wigner Crystallization in quasi-1D nano-wires can be probed with two spatially separated accessible interfaces (two quantum dots) so that nonequilibrium dynamics and its applications can be probed. Particularly we show how this setting can be used to transfer spin qubits between two quantum dots separated by $\mu \text{m}$ scales, which is currently being actively considered as an important problem, with very few suggested solutions~\cite{Trifunovic2012,Trifunovic2013, Mehl2014,Srinivasa2015}. Additionally we suggest a feasible way of observing this phenomenon through pA scale currents, which, in turn, opens up a new option in low dissipation spintronics for a spin current without a charge current.

{\em Wigner Crystalization.-- }By applying strong confining potentials on a 2DEG one may trap a few electrons in a quasi-1D region and effectively make a nano-wire. In such nano-wires, when the electron density is below a critical value, the Coulomb interactions between electrons overtakes their kinetic energies resulting in a quasi-1D Wigner crystal in which the electrons are extremely localised near to the classical equilibrium configurations. In quasi-1D nano-wires where the electrons are strongly confined in two directions, a Wigner crystal is predicted to emerge when the average electron-electron distance is greater than $4a_B$, where $a_B$ is the Bohr radius~\cite{Egger1999}.

\begin{figure*}
\vspace{-100pt}
\begin{center}
	\includegraphics[width=\textwidth]{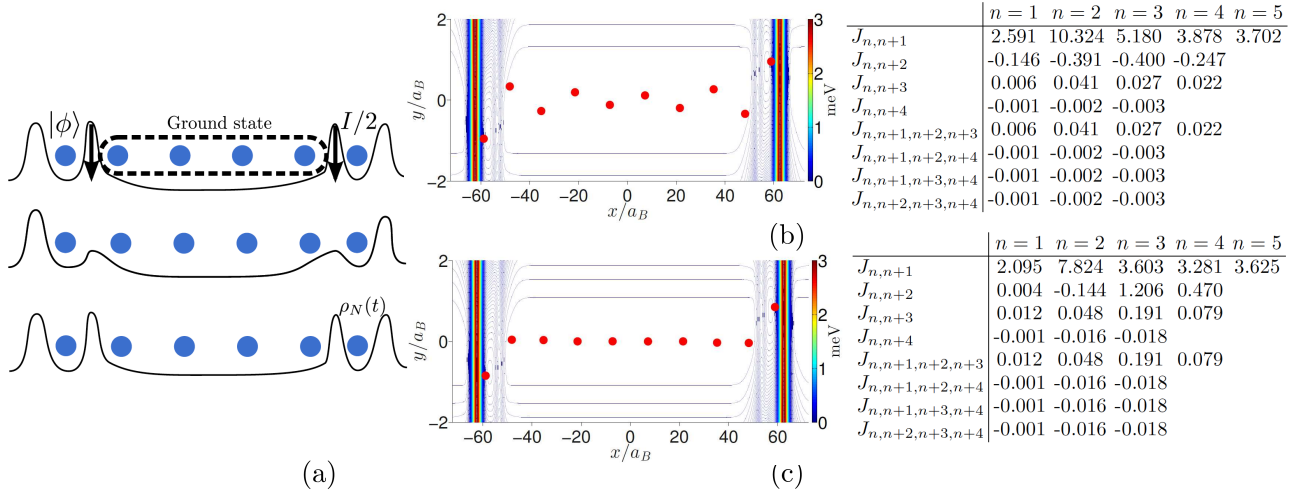}
	\vspace{-100pt}
	\caption{\label{fig:Fig1} (color online) (a) Transfer protocol for an even number of electrons in which $h_b$ changes from $3$meV to $50 \mu$eV instantaneously to induce dynamics.  (b) classical ground state equilibrium positions for 10 electrons with $\Omega =  550\mu eV/\hbar$. Coupling parameters (in the unit of $\mu eV$) are shown for the different 2- and 4-body exchange processes. Since the couplings are symmetric, only half of each coupling is given. (c) Similarly to (b), with $\Omega = 440 \mu eV/\hbar$.}
\end{center}
\end{figure*}

{ \em Model.-- }We consider trapping $N$ electrons (with $N$ even) in a quasi-1D region by using surface electrodes over a 2DEG in GaAs. The trapping potential is modelled as
\begin{eqnarray}\label{Pot_xy}
V_{TR}(x,y)&=&\frac{1}{2}m^*\Omega^2y^2+ h_{0} \left( e^{ - \frac{ (x - d/2)^2 }{ 2 w_{out}^2}} +  e^{ - \frac{(x + d/2)^2 }{ 2 w_{out}^2}} \right)\cr \nonumber
\end{eqnarray}
where $m^*$ is the electron effective mass, $\Omega$ is the strength of the transverse potential and the two Gaussian potentials with height $h_0$ and width $w_{out}$ define a nano-wire of length $d$ extending between $x=\pm d/2$.
Two small quantum dots, for trapping single electrons, are formed at both ends of the wire by applying proper voltages to the gates. The quantum dots are modelled by the following potential
\begin{eqnarray} \label{V_tot}
 V_{QD}(x,y) =  h_{b} \left( e^{ - \frac{(x + d/2 - l)^2 }{ 2 w^2}} + e^{ - \frac{(x - d/2 + l)^2 }{ 2 w^2} } \right) \cr
+ \frac{1}{2}m^* \Lambda^2 \left( e^{-\frac{ (x - x_0)^2}{2 \sigma^2}}   (y - y_0)^2
+  e^{-\frac{(x + x_0)^2}{ 2 \sigma^2}}   (y+y_0)^2 \right)\cr \nonumber
\end{eqnarray}
where the first two Gaussian potentials with heights $h_b$ and width $w$, centered at the distance $l$ from the boundaries of the wire, create two quantum dots at both sides of the nano-wire and the potentials in the second line break the mirror symmetry by displacing the minima in the two quantum dots in opposite directions. The symmetry broken system will have a single equilibrium position for electrons and thus numerical convergence is easier to reach. Choosing appropriate values of the parameters $x_0$, $y_0$ and $\sigma$ is discussed in the supplemental material.

We consider $N=10$ electrons trapped over a distance $d=1.25\mu$m, such that the quantum dots confine one electron each and the remaining 8 electrons are placed in the wire as shown in Fig.~\ref{fig:Fig1}(a). The whole potential is
\begin{eqnarray}\label{V_tot}
  V(\mathbf{R}) &=& \sum_{k=1}^{N} \left[ V_{TR}(\mathbf{r}_k)+V_{QD}(\mathbf{r}_k) +\sum_{j<k} \frac{e^2}{4\pi\epsilon|\mathbf{r}_k-\mathbf{r}_j|} \right]\cr \nonumber
\end{eqnarray}
The barrier $h_b$ is varied from 3meV to $50 \mu$eV to decouple or couple the quantum dots to the wire, respectively. Thus, the two endmost electrons can act as sender and receiver when sending quantum information through the chain. Unlike the middle electrons, we assume full control over these two end electrons for initialization and measurement. 

{\em Exchange couplings.--} At low temperatures the dynamics of the Wigner crystal is approximately governed by the multi-spin-exchange (MSE) Hamiltonian~\cite{Thouless1965}. To calculate the exchange couplings in this Hamiltonian we exploit the semi-classical path integral instanton method~\cite{Roger1983,Hirashima2001,Klironomos2005,Klironomos2007,Meyer2009}.
This approach is applicable for the regimes where the quantum effects are small, like the case for Wigner crystals where the electrons are well separated~\cite{Bernu2001} (see the supplemental material for more details). While our methodology assumes a closed system it is potentially amenable to extension to directly incorporate an environment following the methodology of Ref.~\cite{Reichman-PathIntergral}.

Using the instanton  method, it was found that only processes involving up to $4^{\text{th}}$-nearest neighbour pairwise exchange and up to 4-body exchange were significant. With this restriction, the full spin Hamiltonian can be written (see supplemental material for the details)
\begin{align}\label{eqn:HamSimplified}
H =\sum_{r = 1}^4 \sum_{n=1}^{N-r} J_{n,n+r} \bm{\sigma}_n \cdot \bm{\sigma}_{n+r} + \sum_{j<k<l<m} J_{jklm} \Upsilon_{jklm}
\end{align}
where $\bm{\sigma}_n = (\sigma^x_n ,  \sigma^y_n ,  \sigma^z_n)$ is the Pauli vector acting on site $n$ and $\Upsilon_{jklm} := (\bm{\sigma}_j \cdot \bm{\sigma}_k )(\bm{\sigma}_l \cdot \bm{\sigma}_m ) +  (\bm{\sigma}_j \cdot \bm{\sigma}_m )(\bm{\sigma}_k \cdot \bm{\sigma}_l ) - (\bm{\sigma}_j \cdot \bm{\sigma}_l )(\bm{\sigma}_k \cdot \bm{\sigma}_m )$.
Exchange couplings for $\Omega =  550\mu eV/\hbar$ and $\Omega = 440 \mu eV/\hbar$ are given in units of $\mu eV$ in the tables next to each charge configurations of Figs.~\ref{fig:Fig1} (b) and (c). The general features are low couplings at the boundary (due to the barrier between the dots and the chain) and a U-shaped coupling along the chain. The next-nearest neighbour couplings are always ferromagnetic (i.e. $J_{n,n+2}<0$) for the linear configuration while for the zig-zag geometry they show a more complex pattern varying from negative to positive values.  The U-shaped feature of the couplings is because the off-center electrons are pushed towards the boundaries due to an unbalanced Coulomb repulsion from the majority of electrons on the opposite side. This makes the effective distance between the electrons shorter in the boundaries and thus results in stronger couplings.

{ \em Quantum communication.-- } We assume that the quantum dots are initially decoupled from the wire (i.e.\ $h_b$ is large). Furthermore, we consider zero temperature so that the electrons are confined to their lowest vibrational mode and are prepared in their spin ground state $\ket{GS}$. The electron in one quantum dot is prepared in an arbitrary quantum spin state $\ket{\psi}=\cos(\theta/2) \left| \uparrow \right\rangle +e^{i\phi}\sin{\theta/2}\left| \downarrow \right\rangle$,  which is supposed to be transferred to the opposite dot, which confines an electron in unpolarized mixed state $I/2$. The initial state of the system is thus $\rho(0)=\ket{\psi} \bra{\psi} \otimes \ket{GS}\bra{GS} \otimes I/2$.
The barriers are then simultaneously lowered (i.e.\ $h_b$ is lowered) in order to start a unitary dynamics under the action of the new Hamiltonian. One can compute the density matrix of the last electron $\rho_N(t)$ by tracing out the others. To quantify the quality of transfer one can compute the fidelity $F(t)=\langle \psi|\rho_N(t)|\psi\rangle$ which is independent of $\theta$ and $\phi$ due to the SU(2) symmetry of the Hamiltonian.

\begin{figure}[h!]
	\includegraphics[width=0.5\textwidth]{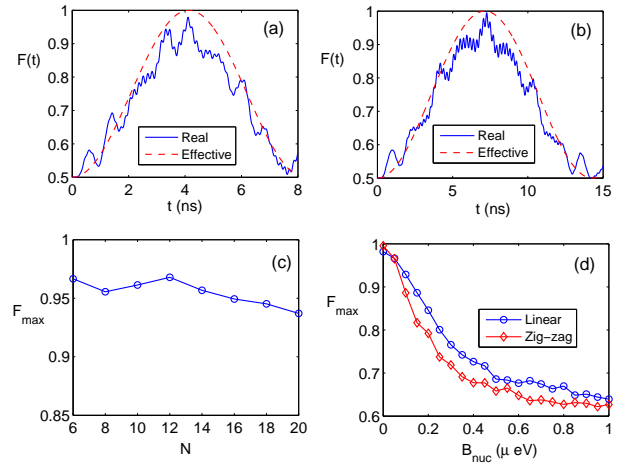}
    \caption{(color online) Average fidelity $F_{av}$ as a function of time, when $h_b$ suddenly changes from 3meV to $50 \mu$eV, using both real and effective Hamiltonians for: (a) the linear chain ($\Omega = 550 \mu eV/\hbar$); and (b) the zig-zag chain ($\Omega = 440 \mu eV/\hbar$). (c) $F_{max}$ versus $N$. (d) The variation in $F_{max}$ when a random magnetic field of variance $B_{nuc}$ is included in the Hamiltonian. }
     \label{fig2}
\end{figure}

In Figs.~\ref{fig2}(a) and (b) we plot $F(t)$ as a function of time for the two charge configurations with $\Omega = 550 \mu eV/\hbar$ (linear chain) and $\Omega = 440 \mu eV/\hbar$ (zig-zag chain). It is clear from these figures that the fidelity peaks at a time $t=t_m$ and the takes its maximum value $F_{max}=F(t_m)$. Although the linear chain gives a faster dynamics with $t_m\simeq 4$ns (due to fairly larger couplings) in comparison with the zig-zag charge configuration with $t_m\simeq 8$ns, the maximum fidelity $F_{max}$ is remarkably high for both configurations, certainly larger than a uniform chain~\cite{Bayat10}. These results illustrate the key point that the details of the charge configurations are not important for the quality of spin transport. To see the scalability, we plot $F_{max}$ as a function of $N$ in Fig.~\ref{fig2}(c), keeping the density of electrons fixed and using only nearest-neighbour couplings since these are by far the largest and computing higher order interactions becomes computationally prohibitive for $>10$ electrons. As the figures shows the fidelity $F_{max}$ remains very high even for chains up to $N=20$ electrons.

{\em Explanation.--} To understand the remarkably high fidelities achieved through the time evolution of the Wigner crystal, one has to consider the nearest-neighbor couplings, which dominate the Hamiltonian $H$. As shown in the tables in Fig.~\ref{fig:Fig1} (b) and (c), the couplings $J_{2,3}$ (which are identical to $J_{N-2,N-1}$) are almost four times stronger than the couplings $J_{1,2}$ (which are identical to $J_{N-1,N}$) in both charge configurations. This implies that the ground state and first excited state of the Hamiltonian $H$ show delocalized strong correlations between the boundary electrons. By computing the reduced density matrix of the two ending spins $\rho_{1N}$ from the ground state of $H$ one can see that $\bra{\psi^-}\rho_{1,N}\ket{\psi^-} >0.8$, where $\ket{\psi^-}$ is the singlet state, for both charge configurations. Similarly, by computing $\rho_{1N}$ from the first excited state of $H$ one gets $\bra{\psi^+}\rho_{1,N}\ket{\psi^+}>0.9$, where $\ket{\psi^+}$ is the triplet state. The delocalized eigenvectors of the Hamiltonian create an effective RKKY-like interaction between the two boundary electrons in the dots~\cite{Venuti2007,Friesen2007}, namely $H_{eff}=J_{eff} \bm{\sigma}_1 \cdot \bm{\sigma}_{N}$ where $J_{eff}=\Delta E/4$ in which $\Delta E$ stands for the energy gap of the Hamiltonian $H$. In Figs.~\ref{fig2}(a)-(b) the time evolution of average fidelities using both $H_{eff}$ and $H$ are plotted which shows that the effective Hamiltonian can qualitatively explain the real dynamics of the system.  In fact, the effective model becomes more precise by decreasing the boundary couplings $J_{12}$ and $J_{N-1,N}$.

{ \em Imperfections.-- }
So far, we have assumed that the system operates at zero temperature and thus the electrons in the wire are initialized in their ground state. In order to guarantee that the proposed protocol remains valid at finite temperature $T$, one has to satisfy $k_B T<\Delta E$, where $k_B$ is the Boltzmann constant. For the given set of couplings the energy gaps are $\Delta E\simeq 0.77$ $\mu$eV for the linear and $\Delta E\simeq 0.45$ $\mu$eV for the zig-zag configurations giving the range of temperature as $k_BT\sim 5-10$ mK which can be achieved in current dilution refrigerators~\cite{batey2013microkelvin}.

In GaAs hetero-structures the electron spins interact with the nuclear spins of the host material. Due to the very slow dynamics of nuclei spins in comparison to the time scales of our protocol one can describe their average effect on electron spin $n$ as an effective random magnetic field $\hat{\textbf{B}}_{n}$. While the direction of this field is fully random its amplitude has a Gaussian distribution~\cite{Taylor2007}
\begin{equation}\label{bnuc}
P(\hat{\textbf{B}})=\exp[- \hat{\textbf{B}}\cdot\hat{\textbf{B}} / 2B_{nuc}^{2} ] / (2\pi B_{nuc}^{2})^{3/2},
\end{equation}
in which $3B_{nuc}^2$ is the variance of the distribution. The total Hamiltonian thus changes as
\begin{equation}\label{H_tot}
H\rightarrow H+\sum\nolimits_{n=1}^N \hat{\textbf{B}}_{n}.\bm{\sigma}_k.
\end{equation}
In Fig.~\ref{fig2}(d) we plot the maximum fidelity $F_{max}$ versus $B_{nuc}$ which shows the destructive effect of the hyperfine interaction. The linear configuration  performs better for larger values of $B_{nuc}$ since the faster dynamics reduces the time exposed to nuclear spins.
A realistic value for $B_{nuc}$ is $2-6$ mT (i.e. $\sim 0.07 - 0.23 \mu \text{eV}$)~\cite{Johnson2005}, at which the fidelity of $F_{max}\simeq 0.8-0.95$ is attainable. Using spin-orbit coupling one may effectively suppress $B_{nuc}$ to $35$ neV \cite{Yacoby-hyperfine-SO}.

At experimental temperatures ($\sim 50 mK$) thermal phonons  are absent in the material \cite{McClintock-1992}.
Moreover, the bulk phonon wavelengths ($\sim \mu m$) exceed electron wavelengths ($\sim 20$ nm) so much that the electrons do not couple to them either. However, sudden quench in $h_b$ may cause the electrons to jiggle around their positions causing fluctuations in exchange interactions. As shown in the supplemental material, these high frequency vibrations ($f \simeq 100 \text{GHz}$) are integrated out during the transmission time $t_m \simeq 5 \text{ns}$.

\begin{figure}

    \includegraphics[width=0.5\textwidth]{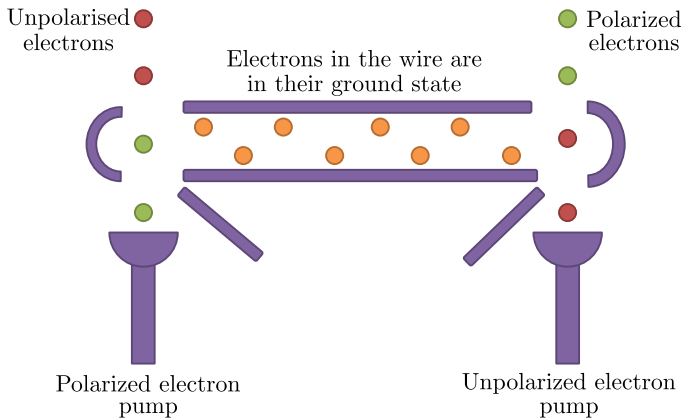}
    \caption{(color online) Schematic diagram of a possible physical implementation of the mechanism in GaAs. Polarised electrons (green circles) are pumped into the left quantum dot, whilst unpolarised electrons (red circles) are pumped into the right dot. Interaction with the confined electrons in the wire (orange circles) swaps the electron spins and thus exchanges the polarization of the currents. A spin polarisation measurement on the output current can detect the transfer of polarised electrons.}
     \label{fig3}
\end{figure}

{\em Experimental Realization.--}  The quantum dot-quantum wire-quantum dot system can be realized in a III-V material such as GaAs/AlGaAs heterostructure~\cite{Kosaka2001}, using gates and quantum point contacts technique, as shown in Fig.~\ref{fig3}. We assume that the quantum wire has achieved a state of Wigner crystallisation~\cite{Meyer_Matveev2007}. Spin initialization in the left dot and read-out in the right dot after the transfer time, can be performed using demonstrated techniques~\cite{Johnson2005,Shulman2012} such as singlet-triplet charge measurements by appending double-dots to both ends of the wire. However, here we show how transport measurements can also probe our mechanism by effectively detecting the transfer of a spin current. The prototype of this measurement consists of a left electron pump~\cite{blumenthal2007,nakaoka2005,Kumar_PRB2014} injecting polarised electrons (say, from a standard source~\cite{Kosaka2001,harrell1999})  whereas the right pump injecting unpolarised electrons in the respective quantum dots as shown in Fig.~\ref{fig3}.  As the electrons are injected into the dots they interact with the electrons in the quantum wire as the barrier height is reduced and their quantum state is swapped. Thus upon exiting the dots, the initially polarized electrons will be unpolarized and vice-versa. As the interaction time for spin swap is about $t_m \sim 4$ns ($t_m\sim 8$ns) for the linear (zig-zag) configuration the electrons must stay in the dot for such time. This implies that the pumps have to inject current of $20\text{pA}- 40\text{pA}$ depending on the configuration of electrons in the quantum wire.
A spin polarization of the current carried off from the right dot by the right pump could then be detected by the halving of the conductance through a quantum point contact in series with the right dot.

{\em Discussion and Conclusions.-- }  We have shown that entirely within the remit of gate defined structures, one can use a Wigner crystal in a quantum wire for transferring spin qubits between quantum dots separated by micrometers reaching very high fidelities. Compared to varied proposals for mediating between separated quantum dot spin qubits~\cite{Trifunovic2012,Trifunovic2013,Mehl2014,Srinivasa2015}, the proposed mechanism can have various advantages. For example, it could have higher speed making it more resilient to decoherence, simpler fabrication compared to hybrid systems, and a greater spatial extension over few electron quantum dot mediators. Furthermore, thanks to the self-assembled nature of the Wigner crystal, our scheme  does not require complex electronics to artificially create a regular structure of exchange coupled spins. Our analysis shows the proposed protocol can operate in current dilution refrigerators, and remarkably the details of the charge configuration does not influence the performance of the system, adding additional robustness to the device fabrication. It also opens an avenue in spintronics by facilitating the spatial transport of a spin current without a charge current, and provides a smoking gun for the near unitary nonequilibrium dynamics in many-body solid-state systems.
Alternative realizations of Wigner crystals, in liquid helium \cite{Wigner-Helium}, ion traps \cite{Wigner-IonTraps} and carbon nanotubes \cite{Wigner-CarbonNanoTubes} can also be used for implementing our mechanism.

{\em Acknowledgements.-- } The authors thank Martin Uhrin for illuminating discussions on use of the 2-point steepest descent algorithm. This work was supported by the Engineering and Physical Sciences Research Council (EPSRC), UK. The research leading to these results has received funding from the
 European Research Council under the European Union's Seventh Framework Programme (FP/2007-2013) / ERC Grant Agreement n. 308253.

\newpage


\widetext
\begin{center}
\textbf{\large Supplemental Materials}
\end{center}

\setcounter{figure}{0}
\setcounter{equation}{0}

\renewcommand{\theequation}{S\arabic{equation}}
\renewcommand{\thefigure}{S\arabic{figure}}

\section{1. Trapping potential}
\label{app:TrapPot}

The trapping potential used in this study is
\begin{align}
  V(\mathbf{R}) = \sum_{k=1}^N \left[ V_{TR}(\mathbf{r}_k)+V_{QD}(\mathbf{r}_k) +\sum_{j<k} \frac{e^2}{4\pi\epsilon|\mathbf{r}_k-\mathbf{r}_j|} \right], \nonumber
\end{align}
where $V_{TR}(\mathbf{r}_k)$, $V_{QD}(\mathbf{r}_k)$ are defined in the main text. An illustration of this potential is given in Fig.~\ref{fig:Potential}.

The external barrier heights in $V_{TR}(\mathbf{r}_k)$ were set to be $h_0 = 3$meV and the length of the system set to $d=1.25 \mu$m as discussed in the main text. We choose a value of $y_0$ such that the the degeneracy between the two possible ground states $\psi_1$ and $\psi_2$ is lifted sufficiently with respect to the temperature, where $\psi_1$ is the mirror inversion of $\psi_2$ about the $x$-axis. The probability of excitation from $\psi_1$ to $\psi_2$ at temperature $T$ in the presence of the symmetry breaking potential $\frac{1}{2}m^* \Lambda^2 (y-y_0)^2$ is roughly $\varepsilon \sim \exp \left(- m^*\Lambda^2 (2y_0)^2 / 2k_B T \right)$. We assume that the system is realised in GaAs, at a temperature of $T\simeq 5 \text{mK}$, with $\Lambda = 500\mu\text{eV} / \hbar$. Inserting these parameters $\varepsilon \sim \exp( -10^{18} y_0^2)$, so $y_0 = 10$nm is used to give a sufficiently low error and since we would expect this level of precision to be possible in experiments.

\begin{figure}[h]
\begin{center}
\includegraphics[width = 0.48\textwidth]{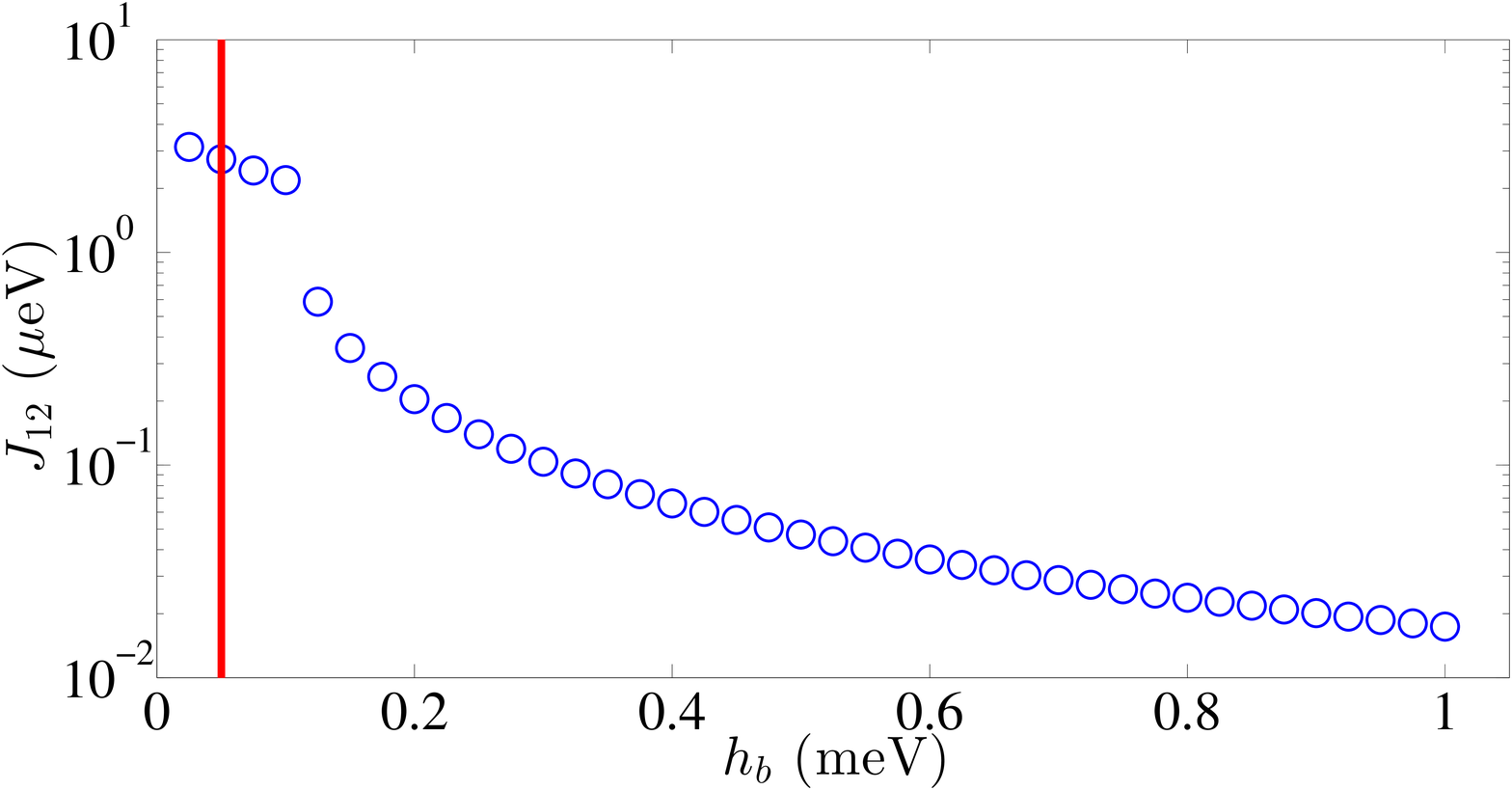}
\caption{\label{fig:BarrierHeights} Variation of the dot-chain coupling $J_{12}$ as the barrier height $h_b$ is changed. The solid red line indicates $h_b = 50\mu \mathrm{eV}$. }
\end{center}
\end{figure}

\begin{figure}[h]
\begin{center}
\includegraphics[width = 0.45\textwidth]{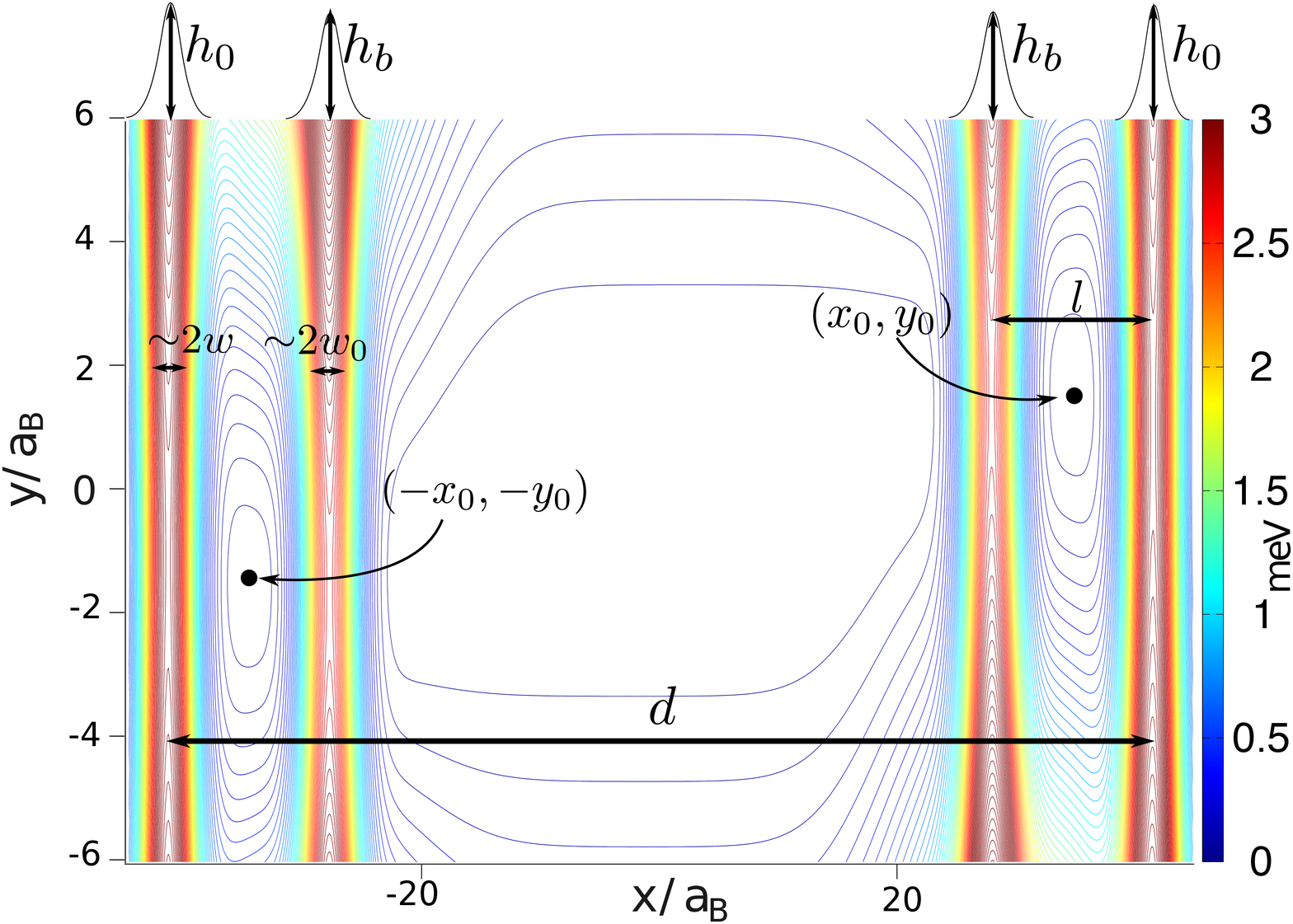}
\caption{\label{fig:Potential} An illustration of the potential $V(\mathbf{R})$ described in the text.}
\end{center}
\end{figure}

The internal barrier height $h_b$ was chosen so that the coupling between the dot electrons and the chain was significant, but the change in the positions of the electrons was still small. A change in $h_b$ from 3meV to 50$\mu$eV was found to change the classical equilibrium electron positions by around $0.1\%$ of average electron separation, whilst still giving a reasonable coupling. The variation in dot-chain coupling for differing barrier heights can be seen in Fig.~\ref{fig:BarrierHeights}.

The remaining parameters are chosen based on the electron configuration without the dot potentials present (i.e.\ when $h_b = \Lambda = 0$). For each particular situation, the minimum energy configuration $\bar{ \mathbf{R}} = (\bar{x}_1,\bar{y}_1,\bar{x}_2,\bar{y}_2,...,\bar{x}_N,\bar{y}_N)$ was found without the dot potentials, and then the parameters were chosen such that the internal barriers would sit roughly halfway between the first two electrons, and such that the internal barriers have as little effect as possible on the separation between the dot electron and the chain. Thus $l = d/2 - (\bar{x}_1 + \bar{x}_2)/2$, and empirically $w = l/8$, $w_{out} =l/8$ was found to have little effect on the electron equilibrium positions. We set $\sigma = l/2$, so that the effect of the two quadratic potentials for the dots is confined to within the barriers. We set $\Lambda =  2\text{meV} / \hbar$ so that the confinement of the dots is much stronger than the confinement in the wire, and to further reduce the susceptibility to thermal noise. The final exchange couplings that we have calculated for this model are well within the experimentally achievable regimes \cite{Shulman2012} which justify the choice of above parameters.

\section{2. Path-integral instanton method}
\label{app:Action}

Before discussing the path-integral instanton method, it will be useful to introduce a characteristic length scale $r_0$, to ease the notation. Following~\cite{Klironomos2007}, we define this as the length scale $r_0 = ( e^2 / 4 \pi \epsilon  m^* \Omega_0^2 )^{1/3}$ at which the Coulomb repulsion between electrons is comparable to the potential energy of the parabolic confinement in the $y$-direction for a particular strength of the $y$-potential  $\Omega_0$, chosen as $\Omega_0 = 0.5 \mathrm{meV}/\hbar$ based on the energy spacing in gated quantum dots~\cite{Taylor2007}. It is also useful to define the dimensionless parameter $r_\Omega=r_0/a_B$, where $a_B$ is the Bohr radius, which gives $r_\Omega \simeq 10$ for the chosen value of $\Omega_0$.

For a system of $N$ electrons constrained to move in 2D, we define collective spatial and spin coordinates $\mathbf{R} = (\mathbf{r}_1,\mathbf{r}_2,...,\mathbf{r}_N)=(x_1,y_1,x_2,y_2...,x_N,y_N)$, $\bm{\sigma} = (\bm{\sigma}_1,\bm{\sigma}_2,...,\bm{\sigma}_N)$, and we use $R_j$ to denote the $j^{th}$ element of $\mathbf{R}$, such that $R_{2j-1} = x_j$, $R_{2j} = y_j$. In the Wigner regime these electrons are localised near to $N$ lattice sites $\mathbf{\bar{R}} = (\bar{x}_1,\bar{y}_1,...,\bar{x}_{N},\bar{y}_N )$.

We assume that the conditions are such that the low-energy Hamiltonian has the form in (\ref{eqn:HThouless}). Then using the path integral instanton approximation, the exchange coupling corresponding to a permutation $P$ of the electron labels is approximately given by~\cite{Voelker2001}
\begin{align}\label{eqn:JPfinal}
J_P \simeq \frac{ e^2 }{4 \pi \epsilon a_B} r_\Omega^{-5/4} A_P F[ \mathbf{R}_{cl}] \left( \frac{\eta[\mathbf{R}_{cl}]  }{2 \pi } \right)^{1/2}  e^{-\sqrt{r_\Omega} \eta[\mathbf{R}_{cl}]}.
\end{align}
where $\mathbf{R}_{cl}$ is the path of least action connecting the configurations $\bar{\mathbf{R}}$ and $P\bar{\mathbf{R}}$, $A_l = 2$ for two body exchange and 1 otherwise, and $\eta[\mathbf{R}]$ is the dimensionless Euclidean action defined as
\begin{align}
\eta[ \mathbf{R} ] :=  \int_0^{T_\tau} d\tau \left[ \sum_j \frac{\dot{\mathbf{r}}_j^2}{2}  + \frac{1}{\hbar \sqrt{r_\Omega}} V(\mathbf{R})\right].
\end{align}
$F[ \mathbf{R}_{cl}]$ is the result of quadratic fluctuations about the classical path, and is given as
\begin{align}\label{eqn:Prefactor}
F[ \mathbf{R}_{cl}] :=  \sqrt{  \frac{ \det( -\partial_{\tau}^2 \idop + \frac{1}{\hbar \sqrt{r_\Omega}}\mathbf{H}^{(0)}(\tau) )  }{ \det'( - \partial_{\tau}^2 \idop + \frac{1}{\hbar \sqrt{r_\Omega}} \mathbf{H}(\tau) )  } }
\end{align}
where $\mathbf{H}$ is the time-dependent Hessian matrix, defined in terms of the potential $V(\mathbf{R})$ as
\begin{align}
\mathbf{H}_{ij}(t) := \left. \frac{\partial^2 V(\mathbf{R})}{\partial  R_i(t) \partial  R_j(t)} \right|_{\mathbf{R} = \mathbf{R}_{cl}}.
\end{align}

In order to numerically compute the exchange couplings $J_P$ we first find the ``classical" configuration that minimises the potential energy of the electrons. The electrons are initialised evenly spaced along the $x$-axis with a slight zig-zag perturbation in the $y$-direction. The configuration that minimises the potential energy is labelled as $\mathbf{\bar{R}}$. Typical minimum energy configurations in a nano-wire of length $d=12 r_0$ for $N=10$ electrons are shown for two different transverse potential $\Omega = 550 \mu \text{eV}$ and $\Omega = 440 \mu \text{eV}$ in Fig. 1 (b) and (c) in the main manuscript. For large values of $\Omega$ (i.e.\ strong confinements), the electrons are constrained to lie along the $x$-axis, whereas for smaller $\Omega$ (i.e.\ weak confinements) a zig-zag pattern forms.

Once the minimum energy classical configuration $\bar{\mathbf{R}}$ has been found, the classical path of least action for each permutation can be numerically found by discretising the exchange path into $M$ equal time steps $\Delta \tau = T_\tau /M$, so that the action integral becomes a Riemann sum:
\begin{align}
\eta[\mathbf{R}] = \int_0^{T_\tau} d\tau \left[  \sum_j \frac{\dot{\mathbf{r}}_j^2}{2} + \frac{1}{\hbar \sqrt{r_\Omega}}V(\mathbf{R})\right] 
\rightarrow \sum_{m=1}^M \sum_j \frac{1}{\Delta \tau} \frac{ (\mathbf{r}_j(\tau_m) - \mathbf{r}_j(\tau_{m-1}))^2}{2} + \frac{\Delta \tau}{\hbar \sqrt{r_{\Omega}} } V(\mathbf{R}_m)
\end{align}
where $\tau_m = m\Delta \tau$ and $\mathbf{R}_m =(r_1(\tau_m),r_2(\tau_m),...,r_{2N}(\tau_m))$. This can be minimised using a steepest descent algorithm, as this is equivalent to the expression for the energy of a chain of beads linked together by springs.

For our calculations, we used $T_\tau = 30$ and $M=70$, based on the work in~\cite{Katano2000} that shows reasonable convergence with these parameters for 27 electrons in a Wigner crystal. The accuracy of discretising time in this way has also been studied in~\cite{Richardson2011} by comparing numerical results to an analytically solvable system. Here it is found that for $M=64$, $T_\tau = 30$ the error is on the order of 1\%.

Once the classical path $\mathbf{R}_{cl}$ has been found, the prefactor $F[\mathbf{R}_{cl}]$ is calculated by diagonalising the numerator and denominator in eqn.\ (\ref{eqn:Prefactor}) so that the determinant can be found. This matrix is calculated along the classical path, and in discretised form becomes a $2N(M+1) \times 2N(M+1)$ matrix. For example, the discretised form of the numerator in eqn.\ (\ref{eqn:Prefactor}) (excluding the square root) is a matrix $\mathbf{A}$ with elements given by
\begin{align}
(\mathbf{A})_{R_i(\tau_k)R_j(\tau_l)} &= -\frac{M^2}{T_\tau^2} \left( -2\delta_{kl} + \delta_{k,l-1} + \delta_{k,l+1} \right) \delta_{ij}
+  \frac{1}{\hbar\sqrt{r_\Omega}}\mathbf{H}_{ij}(\tau_k)\delta_{kl}
\end{align}

\section{3. Spin operators from permutations}
\label{app:Perm}

The multi-spin exchange (MSE) Hamiltonian is given by~\cite{Thouless1965}
\begin{align}\label{eqn:HThouless}
H =  - \sum\nolimits_{P^\sigma \neq \idop} (-1)^{m_P} P^{\sigma} J_{P}
\end{align}
where $P^{\sigma} $ denotes a permutation operator permuting only the spins of the particles according to the permutation $P$ of the electron labels. $m_P$ is the number of 2-body swaps that $P$ can be decomposed into. $J_P$ is the exchange energy, which depends on the energy splitting between the eigenstates of a permutation $P^\sigma$.

\begin{figure}[h]
\begin{center}
	\includegraphics[width=0.5\textwidth]{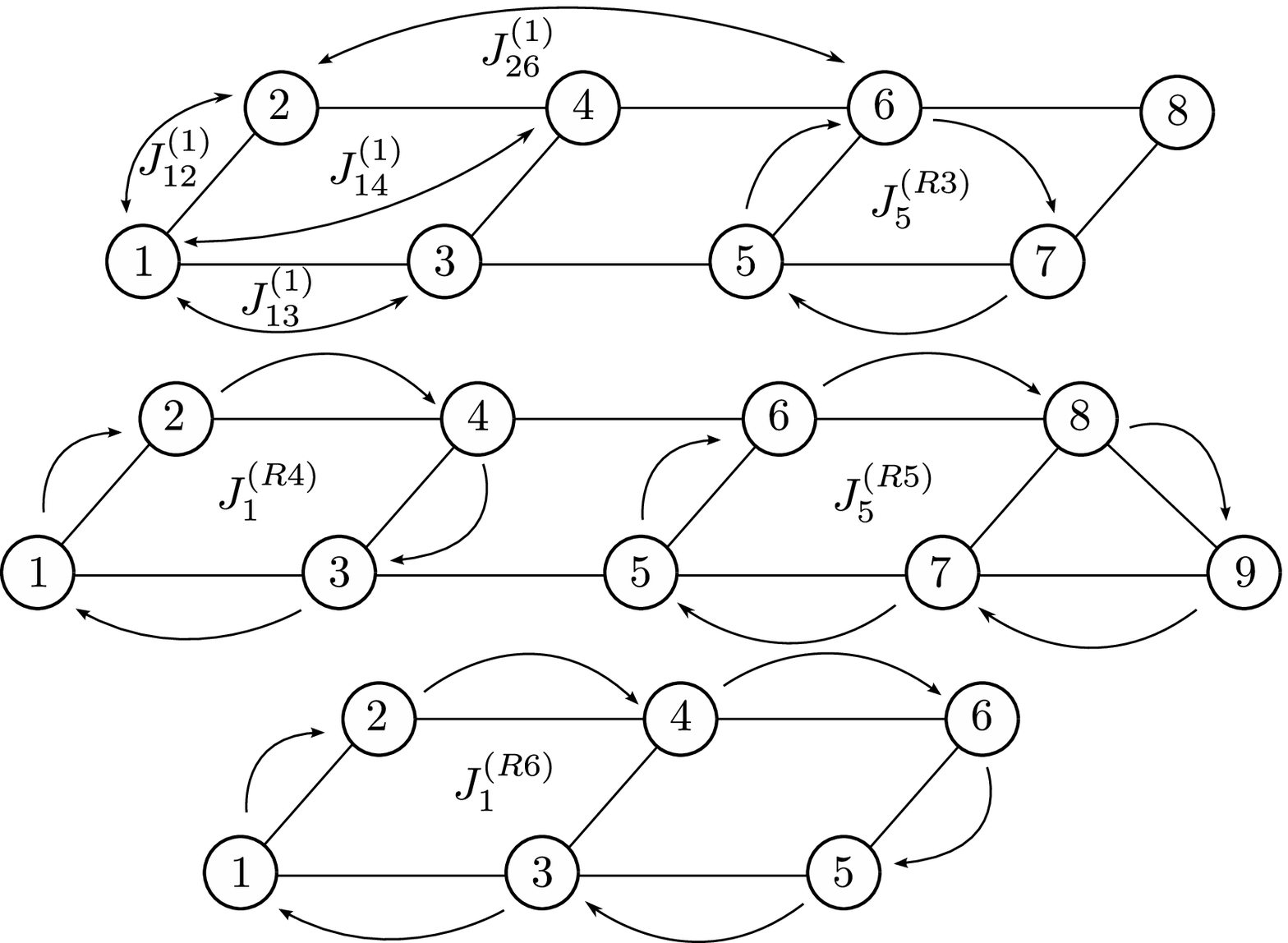}
	\caption{\label{fig:WigExch} An illustration of the types of permutation processes that we consider in this work. $J^{(1)}_{mn}$ indicates a pairwise permutation of electrons $m$ and $n$. $J^{(R3}_{n}, J^{(R4)}_n$, $J^{(R5)}_n$, $J^{(R6)}_n$ indicates 3- 4- 5- and 6-body ring permutations starting clockwise from electron $n$.}
\end{center}
\end{figure}

In this appendix we detail the conversion of the MSE Hamiltonian in (\ref{eqn:HThouless}) into a Hamiltonian with Pauli exchange operators. For the range of parameters considered in this study, permutations of more than 5 electrons and more than nearest neighbour pairwise permutations were found to be 1000 times smaller than the nearest-neighbour exchange couplings over the regime considered here, and so are left out.

With this restriction, the Hamiltonian used to simulate the evolution is
\begin{align}\label{eqn:PermHam}
H = \sum_{m \sim n } J_{mn}^{(2)}  P_{mn}^\sigma  - \sum_{ \substack{ l,m,n \in \triangle \\ l < m < n} } J_{l}^{(R3)} P_{lmn}^\sigma
+ \sum_{ \substack{ k,l,m,n \in \Box \\ k < l< n < m}}  J_{k}^{(R4)} P_{klmn}^\sigma + \sum_{ \substack{ j,k,l,m,n \in \pentagon \\ j < k <n < l<m}} J_{j}^{(R5)} P_{jklmn}^\sigma,
\end{align}
where e.g. $P_{klmn}^{\sigma}$ means a cycle of length 4 which permutes the spins according to $\sigma_k \to \sigma_l, \sigma_l \to \sigma_m, \sigma_m \to \sigma_n, \sigma_n \to \sigma_k$. We use the notation $m \sim n$ to mean $m$ and $n$ are up to 2$^{nd}$-nearest neighbours, $l,m,n \in \triangle$ means that $l,m,n$ are on the vertices of a triangle, $k,l,m,n \in \Box$ means that $k,l,m,n$ are on the vertices of a parallelogram and $j,k,l,m,n \in \pentagon$ means that $j,k,l,m,n$ are on the vertices of trapezium (see Fig.~\ref{fig:WigExch}).

The spin permutation operators can be written in terms of pairwise spin exchange interactions, using the following identities~\cite{Klein1980}:
\begin{align}
&P_{12}^\sigma = \frac{1}{2}(\idop + E_{12}), \\
&P_{123}^\sigma + P_{321}^\sigma = P_{12}^\sigma + P_{23}^\sigma + P_{13}^\sigma - 1,  \\
&P_{1234}^\sigma + P_{4321}^\sigma = P_{12}^\sigma P_{34}^\sigma + P_{14}^\sigma P_{23}^\sigma - P_{13}^\sigma P_{24}^\sigma + P_{13}^\sigma 
+ P_{24}^\sigma -1, \\
&P_{12345}^\sigma + P_{54321}^\sigma =  \frac{1}{2}(P^\sigma_{12}P^\sigma_{34} +  P^\sigma_{14}P^\sigma_{23} -  P^\sigma_{13}P^\sigma_{24})+ \frac{1}{2}(P^\sigma_{15}P^\sigma_{23} +  P^\sigma_{12}P^\sigma_{35} -  P^\sigma_{13}P^\sigma_{25} )\nonumber\\
&+\frac{1}{2}(P^\sigma_{15}P^\sigma_{24} +  P^\sigma_{12}P^\sigma_{45} -  P^\sigma_{14}P^\sigma_{25}) 
 +  \frac{1}{2}(P^\sigma_{15}P^\sigma_{34} +  P^\sigma_{13}P^\sigma_{45} -  P^\sigma_{14}P^\sigma_{35} ) \nonumber\\
&+\frac{1}{2}(P^\sigma_{25}P^\sigma_{34} +  P^\sigma_{24}P^\sigma_{35} -  P^\sigma_{25}P^\sigma_{34}) 
- \frac{1}{2}(P^\sigma_{12}+P^\sigma_{15}+P^\sigma_{23}+P^\sigma_{34}+P^\sigma_{45}) \nonumber\\
&+\frac{1}{2}(P^\sigma_{13}+P^\sigma_{35}+P^\sigma_{25}+P^\sigma_{24}+P^\sigma_{14})- \frac{1}{2}.
\end{align}
By applying the above identities to Hamiltonian (\ref{eqn:PermHam}), this gives the overall Hamiltonian, ignoring the factors of identity since they do not affect the dynamics:
\begin{align}
H &= \frac{1}{2} \sum_{m \sim n }J_{mn}^{(2)}  \bm{\sigma}_m \cdot \bm{\sigma}_n  
-  \frac{1}{2}\sum_{ \substack{ l,m,n \in \triangle \\ l < m < n} } J_{l}^{(R3)}  ( \bm{\sigma}_l \cdot \bm{\sigma}_l  + \bm{\sigma}_m \cdot \bm{\sigma}_n  + \bm{\sigma}_l \cdot \bm{\sigma}_n )  \nonumber\\
&+ \frac{1}{4} \sum_{ \substack{ k,l,m,n \in \Box \\ k < l< n < m}}  J_{k}^{(R4)} \left[  \bm{\sigma}_k \cdot \bm{\sigma}_l  + \bm{\sigma}_k \cdot \bm{\sigma}_m    \right. 
\left.  + \bm{\sigma}_k \cdot \bm{\sigma}_n +\bm{\sigma}_l \cdot \bm{\sigma}_m  + \bm{\sigma}_l \cdot \bm{\sigma}_n  +\bm{\sigma}_m \cdot \bm{\sigma}_n  + \Upsilon_{klmn} \right]  \nonumber\\
&-\frac{1}{8} \sum_{ \substack{ j,k,l,m,n \in \pentagon \\ j < k <n < l<m}} J_{j}^{(R5)} \left[ \Upsilon_{jklm} + \Upsilon_{jkln}  + \Upsilon_{jnkm} + \nonumber\right.
\left. \Upsilon_{jnlm} + \Upsilon_{knlm}
  + \bm{\sigma}_j \cdot \bm{\sigma}_k + \bm{\sigma}_k \cdot \bm{\sigma}_l + \bm{\sigma}_l \cdot \bm{\sigma}_m  \right. \nonumber\\
  & \left. + \bm{\sigma}_m \cdot \bm{\sigma}_n  + \bm{\sigma}_j \cdot \bm{\sigma}_n + \bm{\sigma}_j \cdot \bm{\sigma}_l + \bm{\sigma}_k \cdot \bm{\sigma}_m \right. \left. + \bm{\sigma}_l \cdot \bm{\sigma}_n + \bm{\sigma}_k \cdot \bm{\sigma}_n + \bm{\sigma}_j \cdot \bm{\sigma}_m \right]
\end{align}
where $\Upsilon_{jklm} := (\bm{\sigma}_j \cdot \bm{\sigma}_k )(\bm{\sigma}_l \cdot \bm{\sigma}_m ) +  (\bm{\sigma}_j \cdot \bm{\sigma}_m )(\bm{\sigma}_k \cdot \bm{\sigma}_l ) - (\bm{\sigma}_j \cdot \bm{\sigma}_l )(\bm{\sigma}_k \cdot \bm{\sigma}_m )$. Note the minus sign in front of the $J^{(R3)}$ and $J^{(R5)}$ terms since they involve permutations of an odd number of particles. Overall then, this Hamiltonian can be written
\begin{align}
H =\sum_{r = 1}^4 \sum_{n=1}^{N-r} J_{n,n+r} \bm{\sigma}_n \cdot \bm{\sigma}_{n+r} + \sum_{j<k<l<m} J_{jklm} \Upsilon_{jklm}
\end{align}
where
\begin{align}
J_{n,n+1} &= \frac{1}{2}  J_{n,n+1}^{(2)} - \frac{1}{2}  (J_n^{(R3)} + J_{n-1}^{(R3)})  
+ \frac{1}{4} (J_{n}^{(R4)} +  J_{n-1}^{(R4)} + J_{n-2}^{(R4)}) \cr
&- \frac{1}{8}  (J_{n}^{(R5)} +  J_{n-1}^{(R5)} + J_{n-2}^{(R5)} +  J_{n-3}^{(R5)}))  \\
J_{n,n+2} &= \frac{1}{2}  J_{n,n+2}^{(2)} - \frac{1}{2}  J_n^{(R3)}   + \frac{1}{4} (J_{n}^{(R4)} +  J_{n-1}^{(R4)} )  \cr
&- \frac{1}{8}  (J_{n}^{(R5)} +  J_{n-1}^{(R5)} + J_{n-2}^{(R5)} )  \\
J_{n,n+3} &=  \frac{1}{4} J_{n}^{(R4)} - \frac{1}{8}  (J_{n}^{(R5)} +  J_{n-1}^{(R5)})  \\
J_{n,n+4} &= - \frac{1}{8}  J_{n}^{(R5)} \\
J_{n,n+1,n+2,n+3} &= \frac{1}{4} J_{n}^{(R4)} - \frac{1}{8}  (J_{n}^{(R5)}+ J_{n-1}^{(R5)}) \\
J_{n,n+1,n+2,n+4} &=  - \frac{1}{8} J_{n}^{(R5)}\\
J_{n,n+1,n+3,n+4} &=  - \frac{1}{8} J_{n}^{(R5)}\\
J_{n,n+2,n+3,n+4} &=  - \frac{1}{8} J_{n}^{(R5)}
\end{align}

\section{4. Two-point steepest descent algorithm}
\label{app:SteepDesc}
\begin{figure}[h]
\begin{center}
\includegraphics[width = 0.5\textwidth]{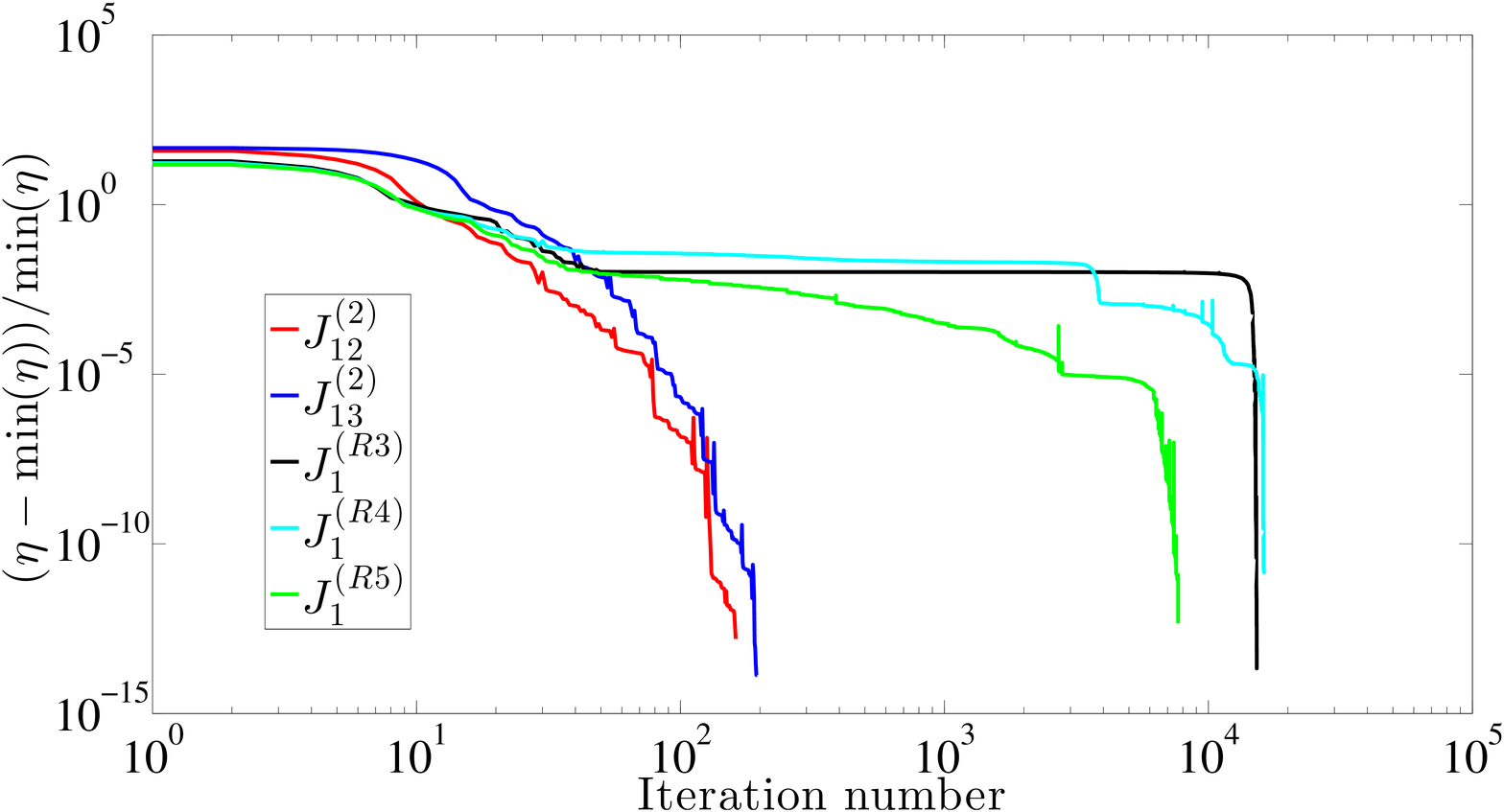}
\caption{\label{fig:Convergence} An example of convergence of the dimensionless action $\eta$ during the 2-point steepest descent algorithm. The trapping potential is $\Omega = 550\mu\text{eV}$, and all processes start from electron 1. }
\end{center}
\end{figure}

The two-point steepest descent algorithm proceeds as follows, with the positions of all of the particles at the $n^{th}$ iteration labelled as $\mathbf{R}_n$~\cite{Barzilai1988}:
\begin{itemize}
\item Make an initial guess $\mathbf{R}_0$.
\item For each step, $\mathbf{R}_{n+1} = \mathbf{R}_n- \alpha_n \mathbf{G}_n$, where $\mathbf{G}_n = \nabla_{ \mathbf{R} }\eta[\mathbf{R}_n] $, $\alpha_0$ is a small step size (we use $\alpha_0 = 0.001$) and $\alpha_n = \mathbf{G}_n \cdot (\mathbf{R}_n - \mathbf{R}_{n-1}) / \|\mathbf{G}_n\|^2$ for $n > 0$~\cite{Barzilai1988}.
\item Stop if $\| \mathbf{G}_n \| < \varepsilon_c$ (convergence to a minimum), or if $\| \mathbf{R}_n \| < \varepsilon_p$ (convergence to a point).
\end{itemize}
Here $\| \cdot \|$ indicates the Euclidean norm, and the tolerance used is $\varepsilon_c = \varepsilon_p = 1\times 10^{-6}$ (although in this study the algorithm always converged to a minimum rather than a point).
A plot showing the convergence of $\eta$ is shown in Fig.~\ref{fig:Convergence}, for all of the exchange processes starting at electron 1 for $\Omega = 550\mu\text{eV}$. The plot shows the change in action relative to the final value $\mathrm{min}(\eta)$, given as $(\eta - \mathrm{min}(\eta))/\mathrm{min}(\eta)$. Clearly the 3, 4 and 5 body processes take much longer to converge than 2-body processes, which is intuitive given the electrons tend to move less for the 2-body processes. The small spikes in the plot can be explained by the algorithm occasionally overshooting the local minimum.

\section{5. Phonon fluctuations}
\label{app:Phonons}
As the barrier is dropped to couple the quantum dot to the nanowire, there may be phonon oscillations induced by the slight change in equilibrium positions. One effect of this would be to cause the exchange couplings to oscillate. To estimate the effects of this on information transfer, we calculated the average change in the nearest-neighbour exchange coupling for different barrier heights. We found that for a change in internal barrier height from $h_b = 3$meV to $h_b=50\mu$eV used in the study, the exchange couplings changed by less than $5\%$. Additionally, we calculated the average classical vibration energy $\bar{\omega}_k = \sqrt{\omega_{x,k}^2 + \omega_{y,k}^2}$ for the $k$th electron where
\begin{align}
\omega_{x,k}^2 = \left. \frac{2}{m}\frac{\partial^2 V}{\partial x_k^2} \right|_{\mathbf{R} = \bar{\mathbf{R}}}, \; \omega_{y,k}^2 = \left. \frac{2}{m}\frac{\partial^2 V}{\partial y_n^2} \right|_{\mathbf{R} = \bar{\mathbf{R}}}
\end{align}
These vibration frequencies are typically on the order of 1-5$\Omega_0 = 0.5- 2.5$meV for the regimes covered in this paper, roughly 100 times larger than the exchange coupling. We can approximately model the effects of these phononic vibrations as adding small random oscillations in the exchange couplings according to
\begin{equation}\label{J_exc_w}
  J \to J[1 + \delta \sin (\bar{\omega} t + \phi)]
\end{equation}
where $\phi$ is a random phase and $\delta$ is a random amplitude chosen from a normal distribution with mean 0 and standard deviation $0.05$. Our numerical simulations show that such fast fluctuation cancels out in time evolution and its impact on the fidelity is negligible. This makes the proposed protocol resistive against phonon excitations created through changing the barrier height $h_b$.

\end{document}